\documentclass{aa}

\usepackage{hyperref}
\usepackage{txfonts}

\newcommand {\Ks} {$K_{\rm s}$}
\newcommand {\mum} {$\muup$m}
\newcommand {\Msun} {$M_\odot$}
\newcommand {\MJup} {$M_{\rm Jup}$}
\newcommand {\Teff} {$T_{\rm eff}$}
\newcommand {\Lbol}{$L_\mathrm{bol}$}
\newcommand {\Lsun}{$L_\odot$}
\newcommand {\logg}{$\log\,g$}
\newcommand{\hd}{HD\,106906}
\newcommand{\hdb}{HD\,106906\,b}

\begin{document}

\title{High signal-to-noise spectral characterization\\of the planetary-mass object HD\,106906\,b
  \thanks{Fully reduced spectra are available in electronic form at the CDS via anonymous ftp to cdsarc.u-strasbg.fr (130.79.128.5) or via \url{http://cdsweb.u-strasbg.fr/cgi-bin/qcat?J/A+A/}.}$^,$\thanks{Based on observations collected at the European Organisation for Astronomical Research in the Southern Hemisphere under ESO programme 094.C-0672(A).}}

\titlerunning{Near-Infrared spectroscopy of HD\,106906\,b}
\authorrunning{Daemgen et al.}

\author{Sebastian Daemgen\inst{\ref{inst1}}
\and Kamen Todorov\inst{\ref{inst2}}
\and Sascha P.\ Quanz\inst{\ref{inst1}}
\and Michael R.\ Meyer\inst{\ref{inst1},\ref{inst3}}
\and Christoph Mordasini\inst{\ref{inst4}}
\and Gabriel-Dominique Marleau\inst{\ref{inst4}}
\and Jonathan J.\ Fortney\inst{\ref{inst5}}
}

\institute{ETH Z\"urich, Institut f\"ur Astronomie, Wolfgang-Pauli-Strasse 27, 8093 Z\"urich, Switzerland, \email{daemgens@phys.ethz.ch}\label{inst1}
  \and Anton Pannekoek Institute for Astronomy, University of Amsterdam, Science Park 904, 1098 XH Amsterdam, Netherlands\label{inst2}
  \and Department of Astronomy, University of Michigan, 1085 S. University, Ann Arbor, MI 48109, USA\label{inst3}
  \and Physikalisches Institut, Universit\"at Bern, Gesellschaftstrasse 6, 3012 Bern, Switzerland\label{inst4}
  \and Department of Astronomy \& Astrophysics, 1156 High Street, University of California, Santa Cruz, CA 95064, USA\label{inst5}
}

\abstract
    {Directly imaged planets are ideal candidates for spectroscopic characterization of their atmospheres. The angular separations that are typically close to their host stars, however, reduce the achievable contrast and thus signal-to-noise ratios (S/N).}
    {We spectroscopically characterize the atmosphere of \hdb, which is a young low-mass companion near the deuterium burning limit. The wide separation from its host star of 7.1\arcsec\ makes it an ideal candidate for high S/N and high-resolution spectroscopy. We aim to derive new constraints on the spectral type, effective temperature, and luminosity of \hdb\  and also to provide a high S/N template spectrum for future characterization of extrasolar planets.}
    {We obtained 1.1--2.5\,\mum\ integral field spectroscopy with the VLT/SINFONI instrument with a spectral resolution of R$\approx$2000--4000. New estimates of the parameters of \hdb\ are derived by analyzing spectral features, comparing the extracted spectra to spectral catalogs of other low-mass objects, and fitting with theoretical isochrones.}
    {We identify several spectral absorption lines that are consistent with a low mass for \hdb. We derive a new spectral type of L1.5\,$\pm$\,1.0, which is one subclass earlier than previous estimates. Through comparison with other young low-mass objects, this translates to a luminosity of log($L$/\Lsun)=$-3.65\pm0.08$ and an effective temperature of \Teff\,=\,1820\,$\pm$\,240\,K. Our new mass estimates range between $M$\,=\,11.9$^{+1.7}_{-0.8}$\,\MJup\ (hot start) and $M$\,=\,14.0$^{+0.2}_{-0.5}$\,\MJup\ (cold start). These limits take into account a possibly finite formation time, i.e., \hdb\ is allowed to be 0--3\,Myr younger than its host star. We exclude accretion onto \hdb\ at rates $\dot{M}$\,$>$\,4.8$\times$$10^{-10}$\,\MJup{}yr$^{-1}$ based on the fact that we observe no hydrogen (Paschen-$\beta$, Brackett-$\gamma$) emission. This is indicative of little or no circumplanetary gas. With our new observations, \hdb\ is the planetary-mass object with one of the highest S/N spectra yet. We make the spectrum available for future comparison with data from existing and next-generation (e.g., ELT and JWST) spectrographs.}
    {}


\maketitle

\section{Introduction}
Direct imaging has revealed more than a dozen planetary-mass companions around young stars \citep[for a review see, e.g.,][]{bow16}. Intermediate resolution ($R$$\gtrsim$1000--2000) infrared spectroscopy of these objects reveals a large number of spectroscopic features that can be compared with those of free-floating objects of similar mass and/or temperature (e.g., brown dwarfs) as well as atmospheric models to constrain formation and early evolution scenarios. Owing to the mostly high contrast ratios and close separations to their host stars, however, only few planets have been studied spectroscopically at high signal-to-noise ratios (S/N). A prime candidate for further investigation is the \hd\ AB+b system. 

\object{HD 106906} is a close binary star \citep{lag17} at a distance of 102.8$\pm$2.5\,pc \citep{gai16} in the Lower Centaurus Crux association \citep[13$\pm$2\,Myr][]{pec12}. This binary star is known to harbor a circumstellar disk extending to $>$500\,AU with a large inner hole and a co-moving low-mass companion at a projected separation of 7\farcs1 \citep[$\sim$730\,AU;][]{che05,bai14,kal15,lag16}. Previous estimates of the temperature and mass of the companion, based on low-resolution 1--2.5\,\mum\ spectroscopy and 0.6--3.5\,\mum\ photometry, placed it in the planetary-mass regime \citep[\Teff=1800\,K, $M$=11$\pm$2\MJup;][]{bai14,wu16}. 

The large distance of \object{HD 106906 b} to its host has spawned discussion about its formation process. This close binary may either have formed like a star through gravitational collapse of a molecular cloud or it may have formed in the primary's disk. The latter may have happened in situ or closer to the star and scattered to its current position through interaction with the central binary and/or other stars \citep[e.g.,][]{rod17}. Alternatively, it may have formed around another star and was scattered \emph{into} the system through a close encounter early in the history of the association \citep{par12}.

We present here new high S/N infrared spectroscopy of \hdb. The observation and data reduction are described in Sect.~\ref{sec:obs}, determinations of spectral type, effective temperature, luminosity, and mass are presented in Sect.~\ref{sec:results}. We summarize the new findings and discuss their implications in Sect.~\ref{sec:summary}.

\section{Observations and data reduction}\label{sec:obs}
Observations of \hdb\ were taken with the SINFONI integral field spectrograph on the Very Large Telescope between December 2014 and March 2015. The primary star \hd\ ($V$=7.8\,mag), 7\farcs1 from the target, served as natural guide star for the adaptive optics system. We obtained spectra in $J$ (1.10--1.40\,\mum), $H$ (1.45--1.85\,\mum), and \Ks\ (1.95--2.45\,\mum) bands with a spatial pixel scale of 125\,mas\,$\times$\,250\,mas, resulting in a field of view of 8\arcsec$\times$8\arcsec\  and a spectral resolution of $R$$\approx$2000--4000. Object (O) and sky (S) observations followed an OSSOOSS\dots\ pattern with a $\sim$1--2\arcsec\ offset between the two positions. The object frames were randomly offset with respect to each other within a radius of $\sim$1\arcsec, always keeping the bright primary outside the field of view. Reference stars with spectral types between B3 and B9 for the correction of telluric absorption were observed close in time to each science observation with the same instrumental setup and at similar airmass. A summary of the observation details is given in Table~\ref{tab:obs}.

\begin{table*}
  \caption{Observation summary\label{tab:obs}}
\centering\small
\begin{tabular}{lclccccccccc}
\hline\hline
  &
  &
  &
  &
  &
  seeing &
  Strehl &
  FWHM\tablefootmark{c} &
  Telluric &
  SpT &
  \Teff\tablefootmark{d} &
  airmass\\
  UT Date  &
  Filter   &
  $n_{\rm exp}$$\times$$t_{\rm int}$ &
  $R\!=\!\lambda/\Delta\lambda$\tablefootmark{a} &
  airmass &
  (arcsec) &
  ratio\tablefootmark{b} &
  (arcsec) &
  ref. star &
  (ref. star) &
  (ref. star) &
  (ref. star)\\
\hline
2014-12-31 & $H$ & 5$\times$100\,s                  & $\sim$3000 & 1.32 & 1.1 & 10--16   & 0.35 & HIP\,050038 & B5 & 15200 & 1.32 \\  
2015-01-23 & $H$ & 2$\times$100\,s                  & $\sim$3000 & 1.18 & 0.6 & $\sim$38 & 0.32 & HIP\,055480 & B8 & 11400 & 1.59 \\  
2015-02-23 & $K$ & 4$\times$300\,s                  & $\sim$4000 & 1.17 & 1.0 & 20--27   & 0.26 & HIP\,059363 & B9 & 10500 & 1.19 \\
2015-02-26 & $J$ & 3$\times$300\,s\tablefootmark{e} & $\sim$2000 & 1.21 & 1.2 & 5--15    & 0.46 & HIP\,055938 & B3 & 19000 & 1.26 \\  
2015-03-05 & $J$ & 6$\times$300\,s                  & $\sim$2000 & 1.17 & 1.3 & 18--27   & 0.46 & HIP\,057474 & B7 & 12500 & 1.24 \\  
\hline
\end{tabular}
\tablefoot{
  \tablefoottext{a}{From the SINFONI manual.}
  \tablefoottext{b}{According to SINFONI's Real Time Computer (RTC).}
  \tablefoottext{c}{Measured at central wavelength.}
  \tablefoottext{d}{Converted from spectral type using \citet{cox_allens_2000}.}
  \tablefoottext{e}{A total of four spectra were observed and extracted at this epoch, but one showed exceptionally high noise and was excluded from further analysis.}
}
\end{table*}

Data reduction partly relies on the SINFONI pipeline (Ver.~2.7.0) in the esorex\footnote{\url{http://www.eso.org/sci/software/cpl/esorex.html}} environment. We used the standard settings and workflow until one integral field data cube was reconstructed per exposure. This involves dark subtraction, flat fielding, distortion correction, and wavelength calibration. The sky frame closest in time, reduced in the same way, was subtracted from each science frame. Since sky frames were taken with a small spatial offset relative to the science frames, most of these frames contain an image of the science target close to the edge of the detector. In order to avoid over subtraction, some sky frames had to be rejected if their point spread function (PSF) overlapped with the science target. Subsequently, we replaced any pixels flagged by the pipeline as bad by an interpolation of the nearest 6 good pixels in spectral direction using a custom \emph{IDL} routine. The resulting integral field spectroscopic cubes have 64$\times$64 pixels\footnote{The SINFONI image plane is filled with 32$\times$64 rectangular pixels. To create the cube, every spatial pixel was automatically split by the pipeline into two neighboring square pixels with identical pixel values.} along the spatial and 2202 pixels along the spectral direction. 

We extracted 1D spectra from the 3D cubes with the procedure described in \citet{dae13}, using apertures with radii identical to 0.8 times the full width at half maximum (FWHM) in each wavelength bin. The FWHM values were determined from Gaussian fits to the trace in the high spatial resolution direction of the cube. Larger apertures could not be used because of contamination of the sky frames with astrophysical sources close to the science target location. We investigated whether any systematic uncertainties are introduced by choosing a small extraction aperture. Using the standard star as a reference, we increased the aperture size from 0.8 to 4$\times$FWHM and found variations of the extracted flux of $<$10\% per pixel at all wavelengths. Because of the low S/N of the target object in each individual wavelength bin, determination of the centroid and FWHM as a function of wavelength use a 2D Gaussian fitting to the bright telluric standard star in every slice of the cube. The extracted trace was shifted to the location of the target by fitting a 2D Gaussian profile to the target PSF after averaging 20--100 slices of the cube in spectral dimension.

The telluric standard stars were reduced and extracted in the same way as the science observations, but with a larger aperture (4$\times$FWHM), and divided by a blackbody curve according to their effective temperature (Table~\ref{tab:obs}). Hydrogen absorption lines were replaced by straight lines along the continuum before dividing the science spectra by the tellurics.

\begin{figure*}[tbh]
\includegraphics[width=0.32\textwidth]{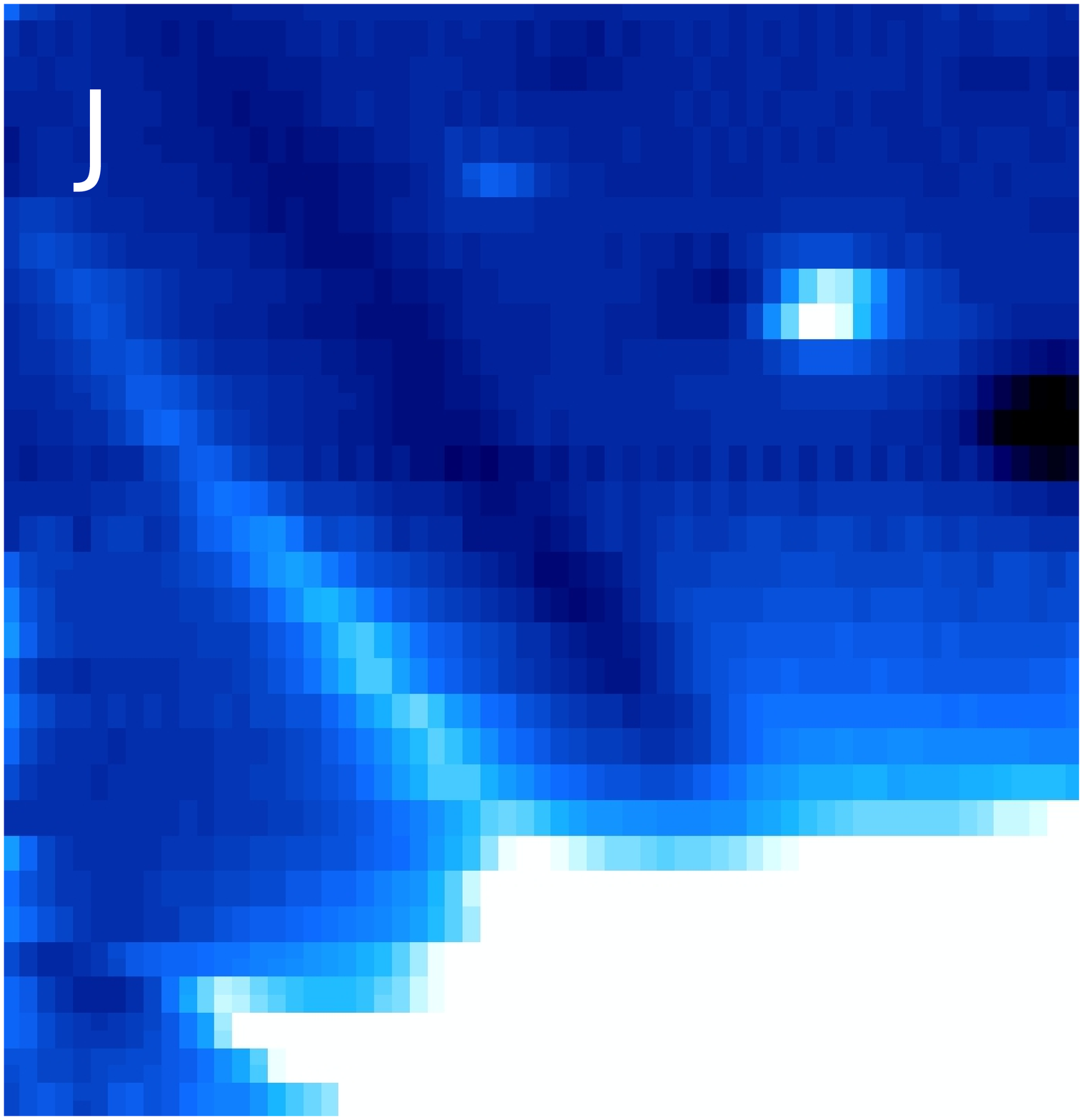}\hfill
\includegraphics[width=0.32\textwidth]{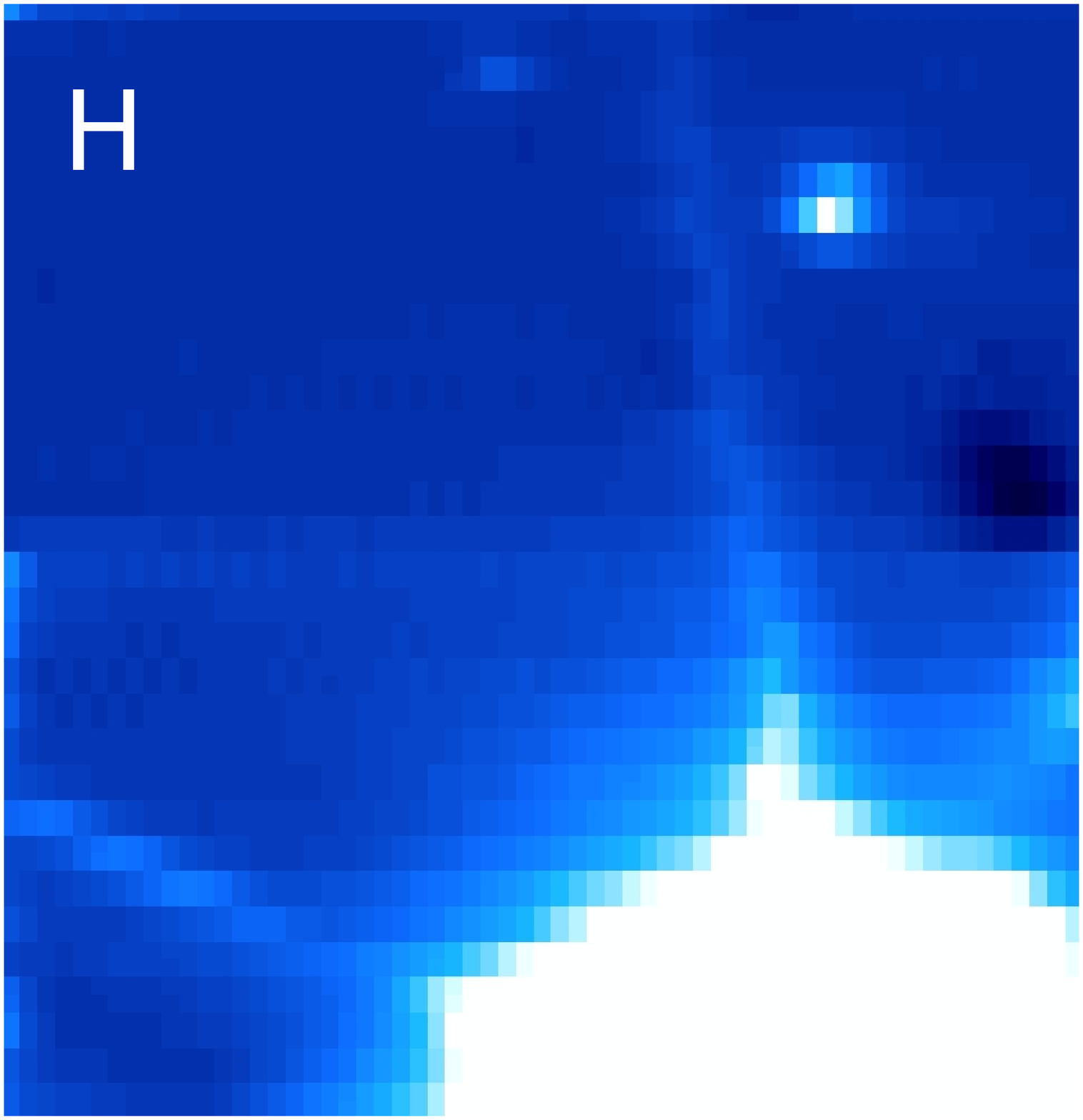}\hfill
\includegraphics[width=0.32\textwidth]{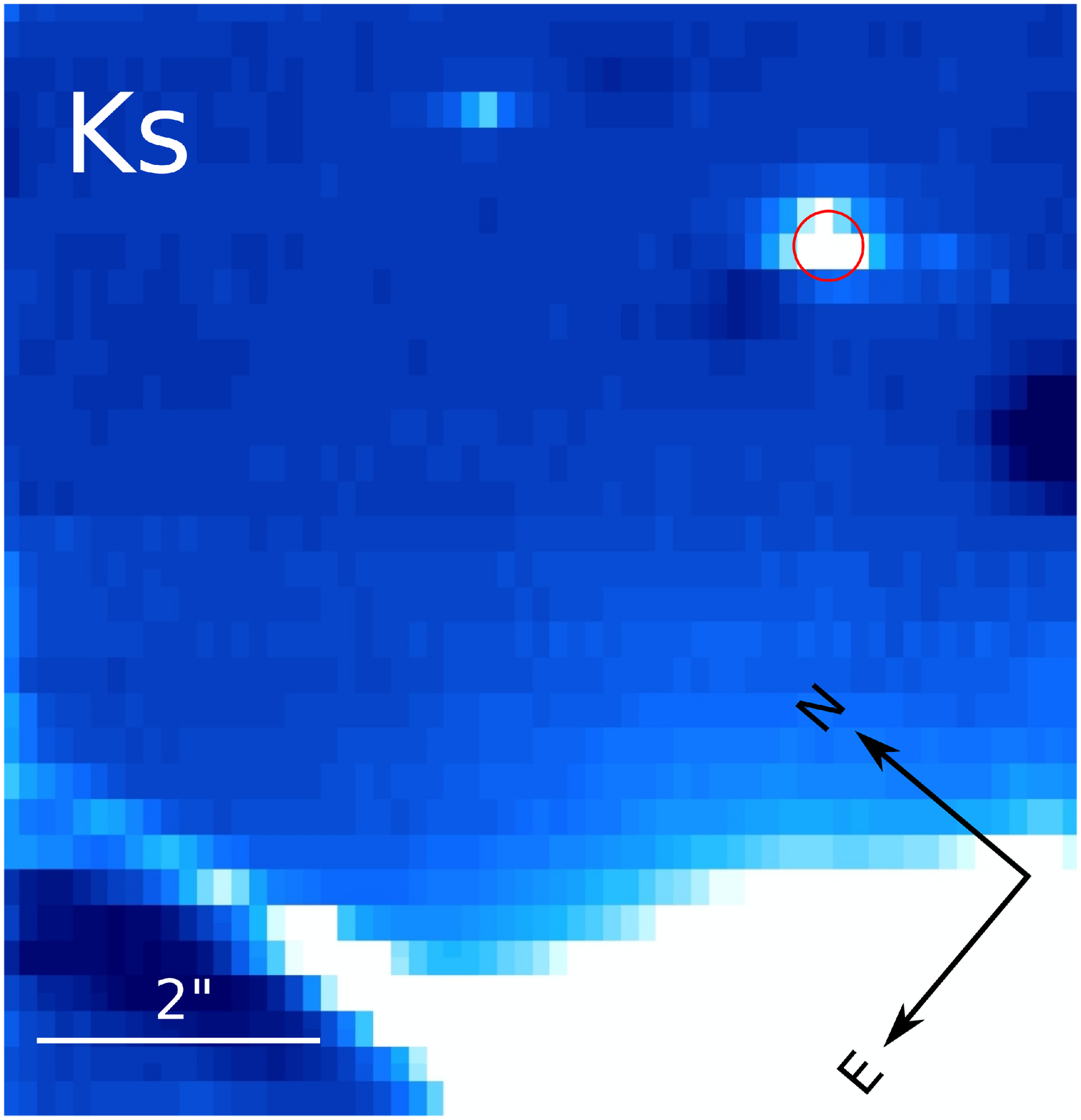}
\caption{\label{fig:im} Examples of individual, fully reduced, and sky-subtracted data cubes of \hdb\ in $J$, $H$, and $K$ band, median-collapsed in spectral direction. The bright region in the bottom right is created by the primary star just outside the field of view. Dark patches are visible where the sky image contains stars. An example aperture for the extraction of the spectra ($r$=0\farcs26) is superimposed on \hdb\ on the $K$-band image in red. Panels show the full field of view of SINFONI (8$\times$8\arcsec); centering is subject to jitter offset. The color stretch is linear, the orientation of all panels as indicated in the \Ks\ image.}
\end{figure*}

\begin{figure*}[tbh]
\centering
\includegraphics[width=0.95\textwidth]{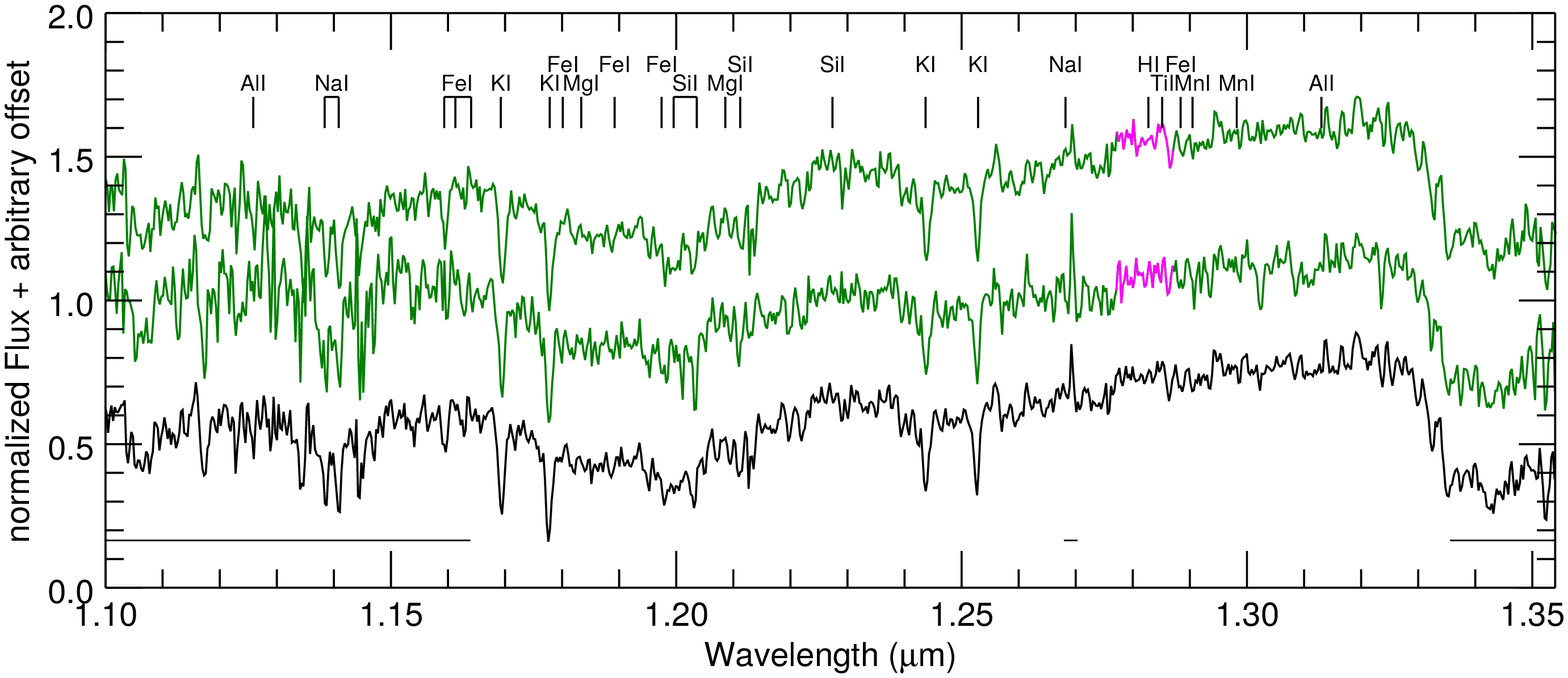}\vspace{1ex}
\includegraphics[width=0.95\textwidth]{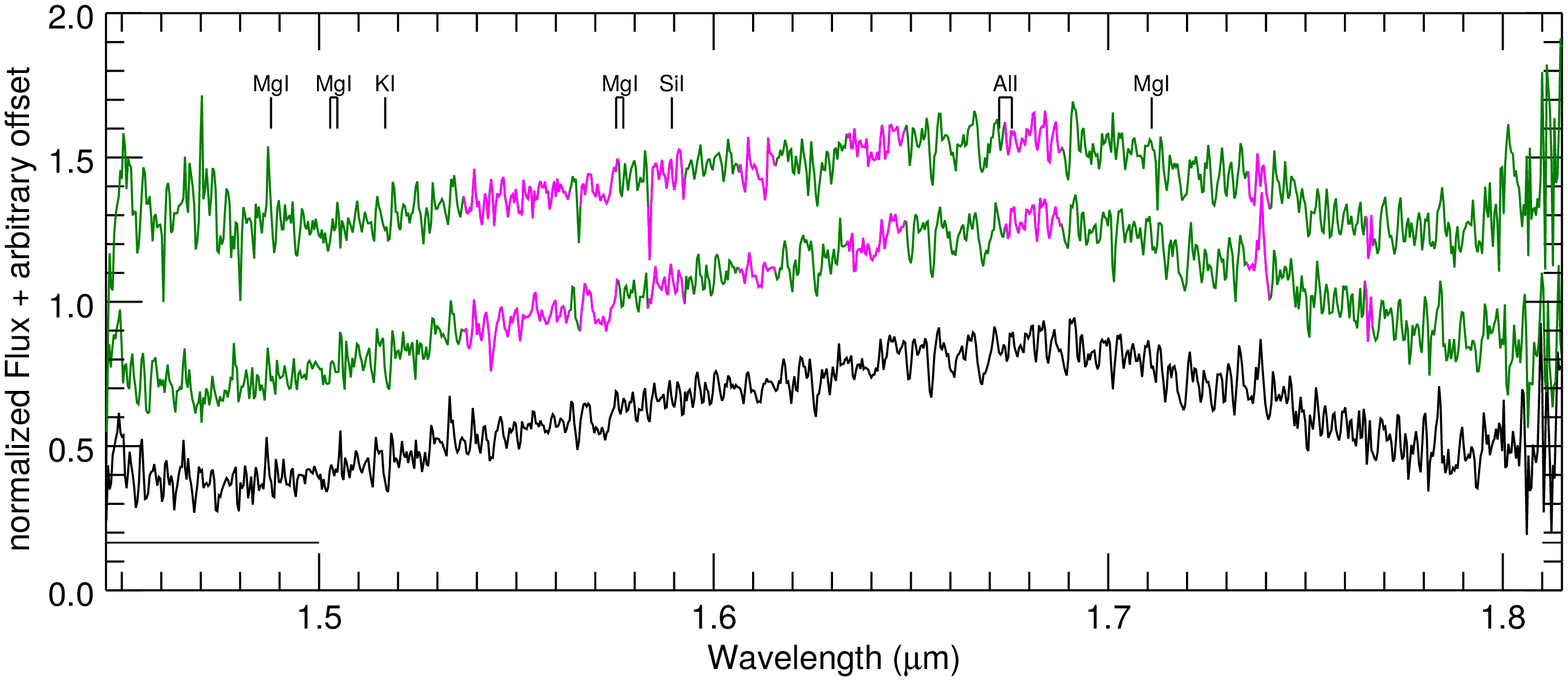}\vspace{1ex}
\includegraphics[width=0.95\textwidth]{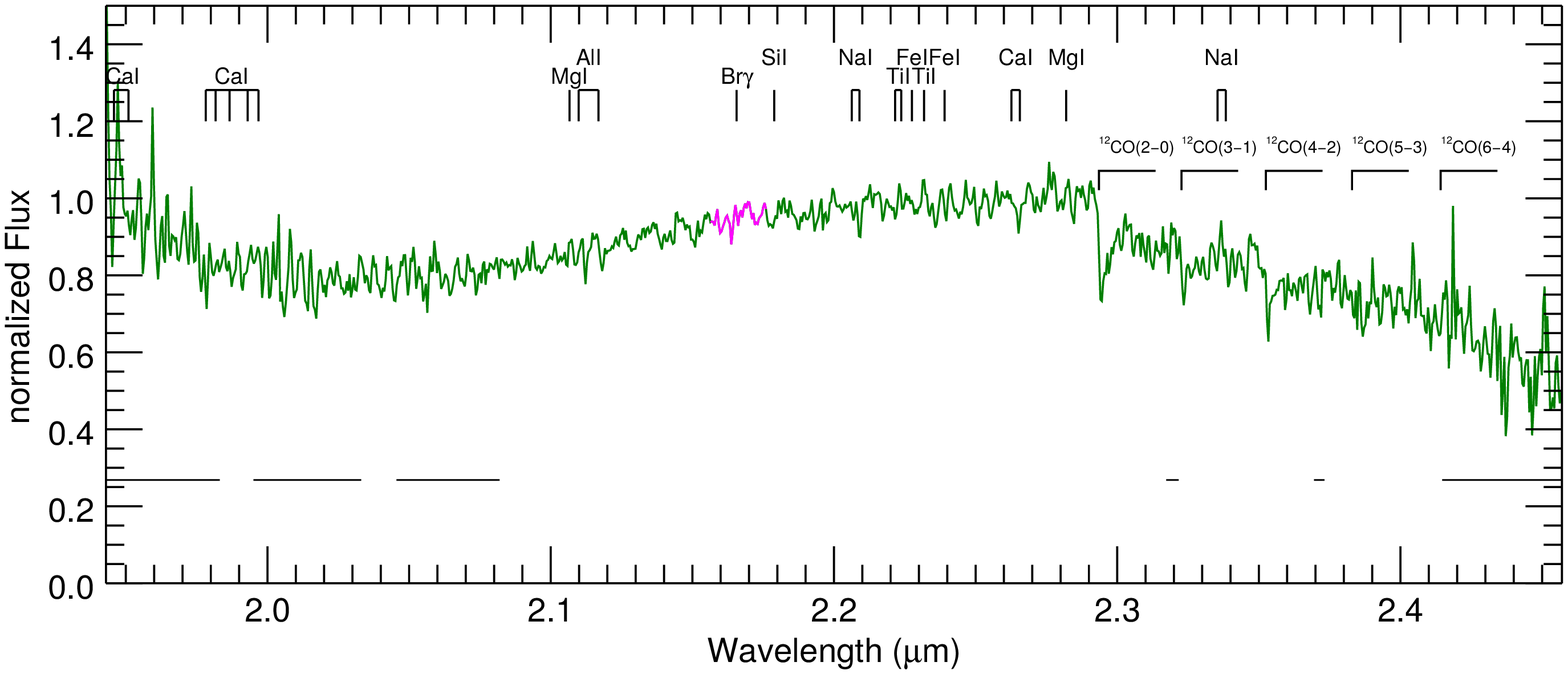}
\caption{\label{fig:spectra}Fully reduced and averaged spectra in $J$ (top), $H$ (middle), and \Ks\ band (bottom) binned by a factor of 2. Green spectra show individual observing periods (most recent on top); black are averaged from all individual exposures per filter. Each spectrum has been normalized individually; offsets and flux scales are arbitrary. Regions highlighted in magenta may contain residual noise from the removal of absorption lines (mostly hydrogen) in the telluric standard. The wavelengths of some atomic and molecular absorption features of low-mass stellar objects are highlighted \citep{cus05}. Regions of strong telluric absorption potentially causing residuals in the reduction are indicated by the black horizontal bars.}
\end{figure*}

Fig.~\ref{fig:im} shows spectral median images of one representative cube per filter. Extracted and averaged spectra are shown in Fig.~\ref{fig:spectra}.
As can be seen in the top two panels of Fig.~\ref{fig:spectra}, we detect slight differences in the spectral slope of the continuum between the two observing epochs in $J$ and $H$ band, respectively. Normalized at the central wavelength in both filters, the amplitude of the variability at the band edges is $\pm$$\sim$10\%. This may be a consequence of the small extraction apertures or of variable telluric water absorption during the observations or a combination thereof. Variable telluric absorption might apply in particular to our second epoch H-band observation where the largest airmass difference between the science and telluric standard target occurred ($\Delta$airmass$\sim$0.4). In addition, intrinsic spectral variability at this level has been previously observed for low-mass objects \citep[e.g.,][]{apa13}. Since we find no significant difference for any of the extracted parameters in Sect.~\ref{sec:results} when analyzing the individual epochs, we average all exposures for each filter band and perform all measurements on the combined spectra.
Uncertainties per spectral wavelength bin are calculated as the standard error of the mean between individual exposures of the same observing sequence. The measured S/Ns are S/N($J$)$\approx$20/pixel, S/N($H$)$\approx$20--50/pix, and S/N($K$)$\approx$20--40/pix.

\section{Analysis and results}\label{sec:results}

\subsection{Spectral characteristics}
The spectra in Fig.~\ref{fig:spectra} show the characteristic triangular shape of the H-band continuum, which is indicative of a young, low-mass, and low surface gravity object. Low gravity is quantitatively supported in the classification scheme by \citet{all13}, which uses ``gravity scores'' based on spectral indices. Scores of 0, 1, and 2 identify objects with no, normal, and strong indications of low gravity, respectively. We measured FeH$_J$=1.12$^{+0.04}_{-0.03}$, K\,I$_J$=1.06$\pm$0.01, and H-cont=1.00$\pm$0.02 resulting in gravity scores of 1, 1, and 2.

We see a large number of spectroscopic features that are likely caused by absorption through atoms and simple molecules in the atmosphere of \hdb. The most prominent features are strong potassium lines between 1.15\,\mum\ and 1.3\,\mum\ as well as clearly detected carbon monoxide bands at $\gtrsim$2.3\,\mum\ \citep{cus05}. These are common features of low-mass objects. Weaker, but significantly detected features include sodium, magnesium, and calcium. The rest of the spectrum appears to be dominated by weaker absorption lines that are partly blended. 

The equivalent widths $W_\lambda=\int (F_\lambda-F_\mathrm{c})/F_\mathrm{c} \,\,\mathrm{d}\lambda$ of a selection of strong ($W_\lambda>1$\,\AA) features are listed in Table~\ref{tab:ews}. 
\begin{table}
  \caption{Equivalent widths of strong ($W_\lambda>1$\,\AA) spectral features\label{tab:ews}}
\centering
\begin{tabular}{ccr@{\dots}lr@{\,$\pm$\,}l}
  \hline\hline
  feature &
  $\lambda_\mathrm{c}$ (\mum) &
  \multicolumn{2}{c}{$\lambda_{\rm int}$ (\mum)\tablefootmark{a}} &
  \multicolumn{2}{c}{$W_\lambda$ (\AA)} \\ 
  \hline
Na\,I          & 1.138 & 1.1360  & 1.1420                    & 7.57 & 0.60 \\
K\,I           & 1.169 & 1.1670  & 1.1710                    & 4.81 & 0.17 \\
K\,I           & 1.177 & 1.1750  & 1.1805                    & 5.81 & 0.39 \\
K\,I           & 1.243 & 1.2415  & 1.2455                    & 5.03 & 0.13 \\
K\,I           & 1.253 & 1.2500  & 1.2550                    & 4.26 & 0.26 \\
K\,I           & 1.520 & 1.5150  & 1.5200                    & 3.64 & 0.29 \\
$^{12}$CO(6$-$3)& 1.618 & 1.6180  & 1.6230                    & 2.63 & 0.16 \\
FeH            & 1.625 & 1.6240  & 1.6280                    & 2.79 & 0.11 \\
FeH?           & 1.650 & 1.6480  & 1.6515                    & 1.76 & 0.09 \\
FeH?           & 1.656 & 1.6540  & 1.6565                    & 2.10 & 0.05 \\
AlI            & 2.110 & 2.1075  & 2.1165                    & 4.11 & 0.40 \\
NaI            & 2.206 & 2.2057  & 2.2105                    & 2.92 & 0.17 \\
CaI            & 2.26  & 2.2610  & 2.2670                    & 2.69 & 0.29 \\
MgI            & 2.281 & 2.2800  & 2.2820                    & 1.03 & 0.08 \\
$^{12}$CO(2$-$0)& 2.294 & 2.2930  & 2.3100                    & 18.6 & 1.1  \\
$^{12}$CO(3$-$1)& 2.323 & 2.3220  & 2.3390                    & 11.6 & 1.1 \\
$^{12}$CO(4$-$2)& 2.353 & 2.3520  & 2.3690                    & 15.3 & 1.7 \\
\hline
\end{tabular}
\tablefoot{
  \tablefoottext{a}{Integration range to determine equivalent widths, taking into account line center shifts with respect to $\lambda_\mathrm{c}$. Line identfication and widths from \citet{mey98,cus05,dae12,bon14}.}
}
\end{table}
The continuum $F_\mathrm{c}$ was measured with a quadratic polynomial fit to visually selected points along the continuum in regions devoid of strong absorption features. Polynomial fits to predefined sections of the spectrum, as used by \citet{bon14}, did not lead to satisfactory fits of the continuum. Uncertainties of $W_\lambda$ were obtained from the standard deviation of repeated measurements with different continuum fits following this recipe.

\subsection{Spectral type}
We determined the spectral type of \hdb\ with three different methods: visual spectral comparisons (Sect.~\ref{sec:vis}), spectral indices (Sect.~\ref{sec:indic}), and examination of the Na\,I and K\,I equivalent widths (Sect.~\ref{sec:ews}). Based on the inferred spectral types of L1.5--L2, L0.9$\pm$0.6, and L0$\pm$2, respectively, we calculated a spectral type of L1.7$\pm$0.3 (inverse variance-weighted average).
This value and its uncertainty is dominated by the visual inspection measurement. Given that we relied on the classifications by \citet{all13} and \citet{bon14}, which themselves are subject to systematic and random uncertainties of typically 0.5 to 1 subclasses, we adopted a conservative uncertainty of 1 subclass for our best estimate. In the commonly used scheme of half subclasses we obtain L1.5$\pm$1.0, which is our best estimate for the spectral type of \hdb. 

\subsubsection{Visual inspection}\label{sec:vis}
We compared our $JHK$ data to the spectral libraries from \citet{all13} and \citet{bon14}, which present and characterize spectra of young late-type objects of low gravity. \citet{all13} examined 73 ultracool young (10--300\,Myr) field dwarfs and classified these sources in a spectral type range of M5--L7. Their spectra have low ($R\approx100$) and medium resolution ($R\gtrsim750$--2000). \citet{bon14} focused on a smaller sample of 15 young brown dwarfs mostly in the M6--L0 range. Their spectra were observed with the same instrument as used for our study (SINFONI) at equal spectral resolution in $J$ band (R$\approx$2000) but slightly lower resolution in $H$ and $K$ ($R$$\approx$1500).

\begin{figure*}[tbh]
  \centering
\includegraphics[width=0.5\textheight]{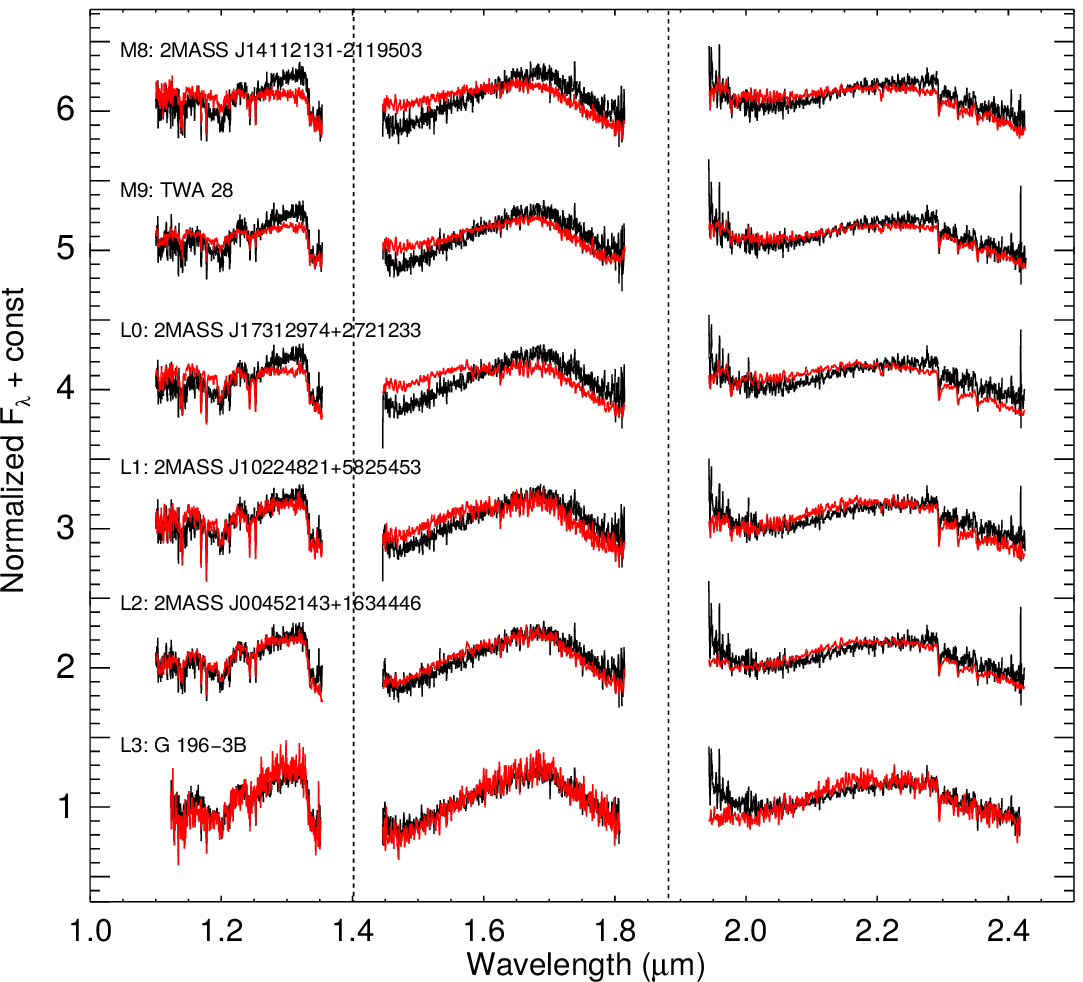}\vspace{1ex}
\includegraphics[width=0.5\textheight]{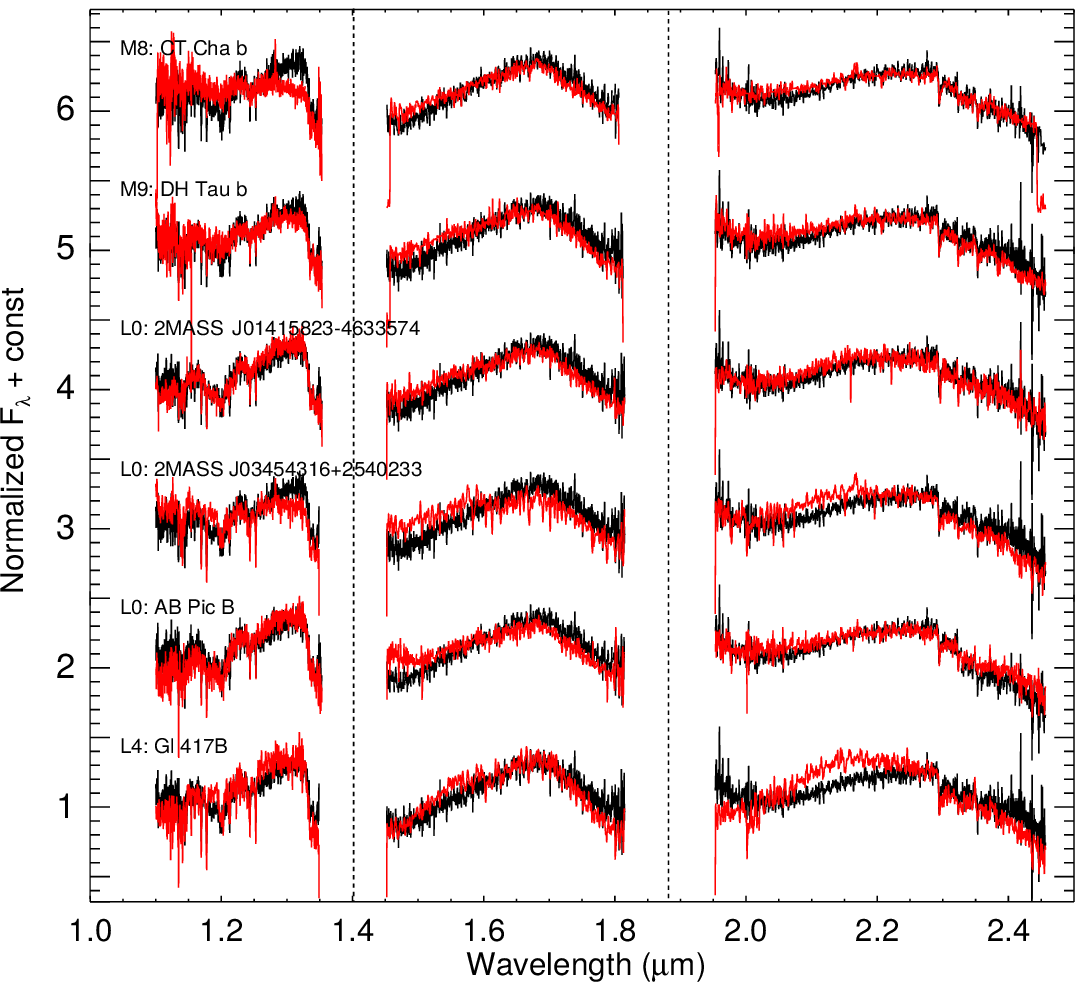}
\caption{\label{fig:visualinspection}Comparison of our spectra (black) with the libraries of young star spectra (red) by \citet[top]{all13} and \citet[bottom]{bon14}. The $J$, $H$, and $K$ spectra have been normalized and matched for each band individually.}
\end{figure*}

In Fig.~\ref{fig:visualinspection} we compare our observed spectra to the library spectra. Because our spectra were not observed simultaneously and we cannot calibrate the flux of the H-band spectrum owing to an unknown H-band magnitude, we cannot obtain a reliable relative flux calibration of the full $JHK$ spectrum. We thus compare each filter individually. 
For each comparison, our target spectrum has been convolved with a Gaussian profile to match the spectral resolution of the respective library spectrum. The individual results are listed in Table~\ref{tab:spectralindices}. Best fits range from M8 to L4. Based on the comparison from the \citet{all13} library, we classify the companion as a L1.5--L2, while the \citet{bon14} library suggests M8--L0. However, a caveat is that the latter library does not contain spectra with types L0.5--L3.5.

\begin{table}
  \caption{Spectral type estimation\label{tab:spectralindices}}
\centering
\begin{tabular}{lcc}
\hline\hline
Data set/Index & Ref. & Spectral type \\
\hline
$J$        & 1 & L0.5\,$\pm$\,1.5 \\
$H$        & 1 & L2.5\,$\pm$\,0.5 \\
$K$        & 1 & L1.0\,$\pm$\,1.0 \\
\multicolumn{2}{l}{{\bf Weighted Average}\tablefootmark{a}} &  {\bf L2.1\,$\pm$\,0.4} \\
\hline\\[-1.5ex]
$J$        & 2 & L0.0\,$\pm$\,1.0 \\
$H$        & 2 & L0.0\,$\pm$\,4.0 \\
$K$        & 2 & M9.5\,$\pm$\,1.5 \\
\multicolumn{2}{l}{{\bf Weighted Average}\tablefootmark{a}} &  {\bf M9.9\,$\pm$\,0.8} \\
\hline
FeH        & 3 & $>$L0.6\tablefootmark{b} \\  
H$_{2}$O-1 & 3 & L0.6\,$\pm$\,1.1 \\  
H$_{2}$O   & 4 & L1.1\,$\pm$\,0.8 \\  
H$_{2}$O-2 & 3 & L1.0\,$\pm$\,1.6 \\  
H$_{2}$OD  & 5 & M7.1\,$\pm$\,0.8\tablefootmark{b} \\  
\multicolumn{2}{l}{{\bf Weighted Average}} &   {\bf L0.9\,$\pm$\,0.6} \\
\hline
\end{tabular}
\tablefoot{Spectral types as estimated from visual inspection (top two sections) and spectral indices (bottom).
  \tablefoottext{a}{Average for a spectral library, weighted by the inverse variance.}
  \tablefoottext{b}{These measurements do not enter our final spectral type estimate because the FeH index does not cover the entire range of interest for \hdb\ \citep[valid only for M3--L3;][]{sle04} and H$_2$OD is based on data that are strongly affected by residuals from the telluric correction \citep[1.96--2.08\,\mum;][]{mcl03}.}
  }
\tablebib{(1) \citealt{all13}; (4) \citealt{bon14}; (3) \citealt{sle04}; (4) \citealt{all07}; (5) \citealt{mcl03}.}
\end{table}

\subsubsection{Spectral indices}\label{sec:indic}
In order to be consistent with previous studies, we adopted spectral indices that were used by \citet{all13} and \citet{bon14}. Specifically, we explored the H$_{2}$O index near 1.5\,\mum\ \citep{all07}, H$_{2}$O-1 ($\sim$1.3\,\mum), and \mbox{H$_{2}$O-2} ($\sim$2.1\,\mum) indices \citep{sle04} and the H$_{2}$OD index \citep[$\sim$2.0\,\mum,][]{mcl03}. Since the $J$ band contains many spectral features from various molecules, we also examined the $J$ band FeH spectral index \citep{sle04}, near 1.2\,\mum. The uncertainties of the best inferred value of L2.5$\pm$1.9, however, extend beyond the valid spectral type range of M3--L3. Thus, we did not include these measurements in our final spectral type estimate.

Table~\ref{tab:spectralindices} summarizes our results for each spectral index. Uncertainties were propagated from our measurement uncertainty and the quoted uncertainties of the polynomial fits in the original publications. 
Calculating the average of the spectral types determined using the H$_{2}$O, H$_{2}$O-1 and H$_{2}$O-2 indices and weighing the average by the inverse of the square of their uncertainty, we get a spectral type of L0.9$\pm$0.6. 

\subsubsection{Equivalent widths}\label{sec:ews}
Our spectra have a high S/N and a relatively high resolution. We can use the additional information contained in our data compared to, for example, \citet{all13} to improve our spectral type determination. We compare the equivalent widths of five $J$-band features of Na\,I (1.138\,\mum) and K\,I (1.169, 1.177, 1.243, and 1.253\,\mum; see Table~\ref{tab:ews}) with those of other young objects as measured by \citet{bon14} at the same spectral resolution. We find that the equivalent widths are broadly consistent with an L0$\pm$2 object. However, the correlation of spectral type with the Na\,I and K\,I equivalent widths of young object spectra such as those of \hdb\ is weak and has a larger scatter compared to field objects, as is apparent from Fig.~11 in \citet{bon14}. This diagnostic thus leads to less precise results than for field dwarfs. More independent determinations of precise equivalent widths and spectral types are needed to increase the usefulness of equivalent widths for spectral type determinations of young planetary-mass objects.

\subsection{Effective temperature and luminosity}\label{sec:teff}
To obtain robust estimates of \Teff\ and \Lbol, we use two strategies. The first determination is based on our best spectral type estimate and uses empirical correlations of \Teff\ and \Lbol\ with spectral type for young late-type objects. The second approach derives a luminosity of \hdb\ based on its brightness, empirically calibrated bolometric corrections, and the new Gaia distance, which is 10\% larger than pre-Gaia estimates.

Our analysis is based on the analysis of colors, luminosities, and effective temperatures of 152 young brown dwarfs and directly imaged planets by \citet{fah16}. They have estimated \Lbol\ by integrating the available SEDs for their sample of objects between 0 and 1000 microns. Since the distances to these objects are known, \citeauthor{fah16} have estimated radii using \Lbol\ and model isochrones. Then, based on radius and \Lbol, they have used the Stefan-Boltzmann law to calculate \Teff. We mostly use the results from the \citeauthor{fah16} study as it supersedes the smaller young object sample by \citet{fil15}. The latter study, however, presents an analysis of a comparison sample of 65 field stars as well as bolometric corrections for young low-mass objects that are used in Sect.~\ref{sec:lum}.

\subsubsection{Effective temperature}
\citet{fah16} have provided polynomial fits to the \Lbol\ and \Teff\ correlations with spectral type for three categories of low-gravity objects. First, their ``YNG'' sample contains all spectra with indications of low gravity. Second, the ``YNG2'' contains YNG as a subset, but also includes directly imaged planetary-mass objects. Third, the ``GRP'' sample contains confirmed young moving group targets.
For our best estimate of the spectral type of \hdb\  (L1.5$\pm$1.0), we obtain $T_{\rm eff}^{\rm YNG}$\,=\,1920\,$\pm$\,210\,K, $T_{\rm eff}^{\rm YNG2}$\,=\,1820\,$\pm$\,240\,K, and $T_{\rm eff}^{\rm GRP}$\,=\,1900\,$\pm$\,240\,K. Uncertainties were derived from both the spectral type uncertainty and the rms of the spectral type-\Teff\ relation by \citet{fah16}. The temperatures are in good agreement with each other and with the \citeauthor{bai14} value of \Teff=1800$\pm$100\,K derived from evolutionary models. Since \hdb\ is best described by the YNG2 sample, we adopt \Teff=1820\,$\pm$\,240\,K as our best estimate for its temperature.

This temperature is lower than what is obtained when assuming that \hdb\ is a field object ($T_{\rm eff}^{\rm field}$\,=\,2030\,$\pm$\,180\,K). It is also slightly lower than the field object value derived by \citealt{bai14} of $T_{\rm eff}^{\rm field}$=$1950\pm200$\,K based on their later spectral type of L2.5$\pm$1. This is because similar to young brown dwarfs, planets are redder than field dwarfs in the near-infrared bands \citep[e.g.,][]{fah16}; this is likely explained by high-altitude clouds \citep[e.g.,][]{bow10,cur11,mad11,mar12,ske12}. 

\subsubsection{Luminosity\label{sec:lum}}
Using the polynomial fits to the empirical luminosity-spectral type relations derived by \citet{fah16}, we calculate values of $\log\left(L_\mathrm{bol}^\mathrm{YNG}/L_\odot\right)$\,=\,$-$3.83\,$\pm$\,0.35, $\log\left(L_\mathrm{bol}^\mathrm{YNG2}/L_\odot\right)$\,=\,$-$3.64\,$\pm$\,0.24, and $\log\left(L_\mathrm{bol}^\mathrm{GRP}/L_\odot\right)$\,=\,$-$3.47\,$\pm$\,0.31. Uncertainties, as estimated from a Monte Carlo simulation, take into account the spectral type uncertainty and the rms in $\log L$ \citep{fah16}. Again, we select YNG2 as the most appropriate value and listed the others to illustrate the range of answers depending on the underlying sample. 

For an independent estimate of the luminosity we also use bolometric corrections of young stars for the $J$ band and \Ks\  band, respectively. These are available through polynomial fits to the YNG sample of \citet{fil15}. With bolometric corrections for the \hdb\  $J$ and \Ks\ magnitudes of $BC_J$\,=\,1.54\,$\pm$\,0.28\,mag and $BC_{Ks}$\,=\,3.27\,$\pm$\,0.13\,mag, we derive luminosities of $\log\left(L_\mathrm{bol}^{\mathrm{YNG,BC}_{Ks}}/L_\odot\right)$\,=\,$-$3.57\,$\pm$\,0.05 and $\log\left(L_\mathrm{bol}^{\mathrm{YNG,BC}_J}/L_\odot\right)$\,=\,$-$3.74\,$\pm$\,0.11 calculated as
\begin{equation}
  \log\left(\frac{L_\mathrm{bol}}{L_\odot}\right)=0.4\left(M_\mathrm{bol,\odot}-m+5\log\left(\frac{d}{10\,\mathrm{pc}}\right)-BC\right)\quad,
\end{equation}
using $M_\mathrm{bol,\odot}$=4.74\,mag. Uncertainties were propagated using a Monte Carlo simulation based on the rms values reported by \citet[$\Delta$BC$_{Ks}$=0.126\,mag, $\Delta$BC$_{J}$=0.189\,mag]{fil15} as well as the uncertainties of the distance modulus and our spectral type estimate.
The comparably faint luminosity derived from the $J$-band corrections is due to a red $J$$-$\Ks\ color, compared to atmospheric models and other targets of this spectral type, which was already observed by \citet{wu16}. 

From the three estimates ($L_\mathrm{bol}^\mathrm{YNG2}$, $L_\mathrm{bol}^{\mathrm{YNG,BC}_J}$, and $L_\mathrm{bol}^{\mathrm{YNG,BC}_{Ks}}$) we derive $\log\left(L_\mathrm{bol}/L_\odot\right)$\,=\,$-$3.65\,$\pm$\,0.08 as our best estimate of the bolometric luminosity of \hdb\  using a Monte Carlo simulation to propagate the uncertainties.

\subsection{Mass}\label{sec:mass}
We used stellar evolution models to translate our inferred luminosity to mass.
Model families include DUSTY \citep{cha00}, BT-Settl \citep{bar15}, and the Bern Exoplanet Tracks (BEX). The first two feature dusty atmospheres and are classified as ``hot start'' according to their high initial (i.e., post-formation) entropies. The BEX use initial entropies found in self-consistently coupled planet formation and evolution simulations \citep{mor12}. The models were updated to employ Ames Cond atmospheric boundary conditions \citep{all01} and include Deuterium burning \citep{mol12}. The coupled formation and evolution calculations lead in population syntheses to planets with a range of post-formation entropies and deuterium abundances for more massive objects. This allows us to identify initial conditions for the evolution corresponding to the hottest, hot, warm, and cold post-formation states \citep{mor17}.

\begin{figure*}[tbh]
  \centering
  \setlength{\unitlength}{\textwidth}
\begin{picture}(1.0,1.2)
    \put(0.00,0.80){\includegraphics[width=\columnwidth]{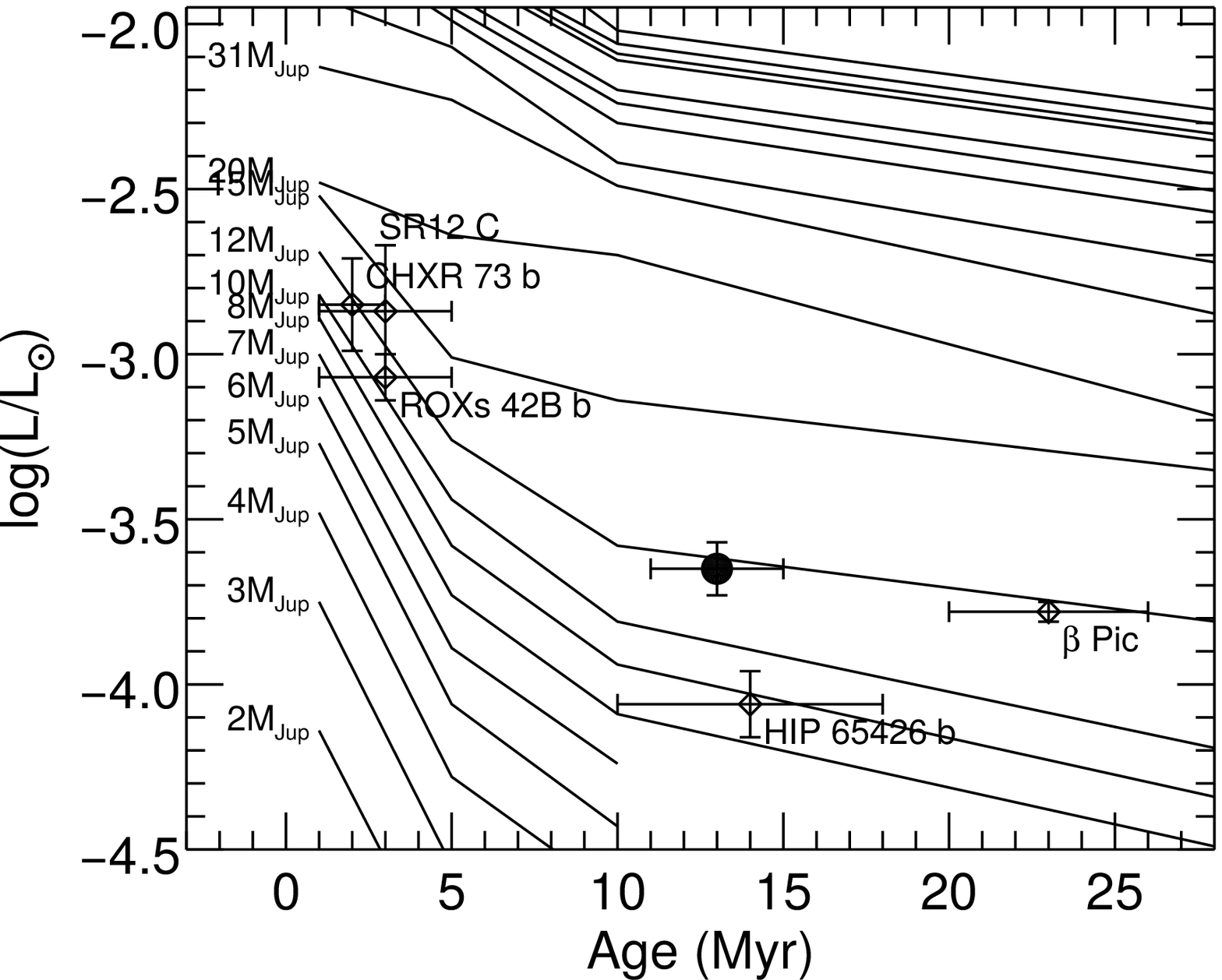}}
    \put(0.40,1.15){\fontfamily{phv}\selectfont DUSTY}
    \put(0.50,0.80){\includegraphics[width=\columnwidth]{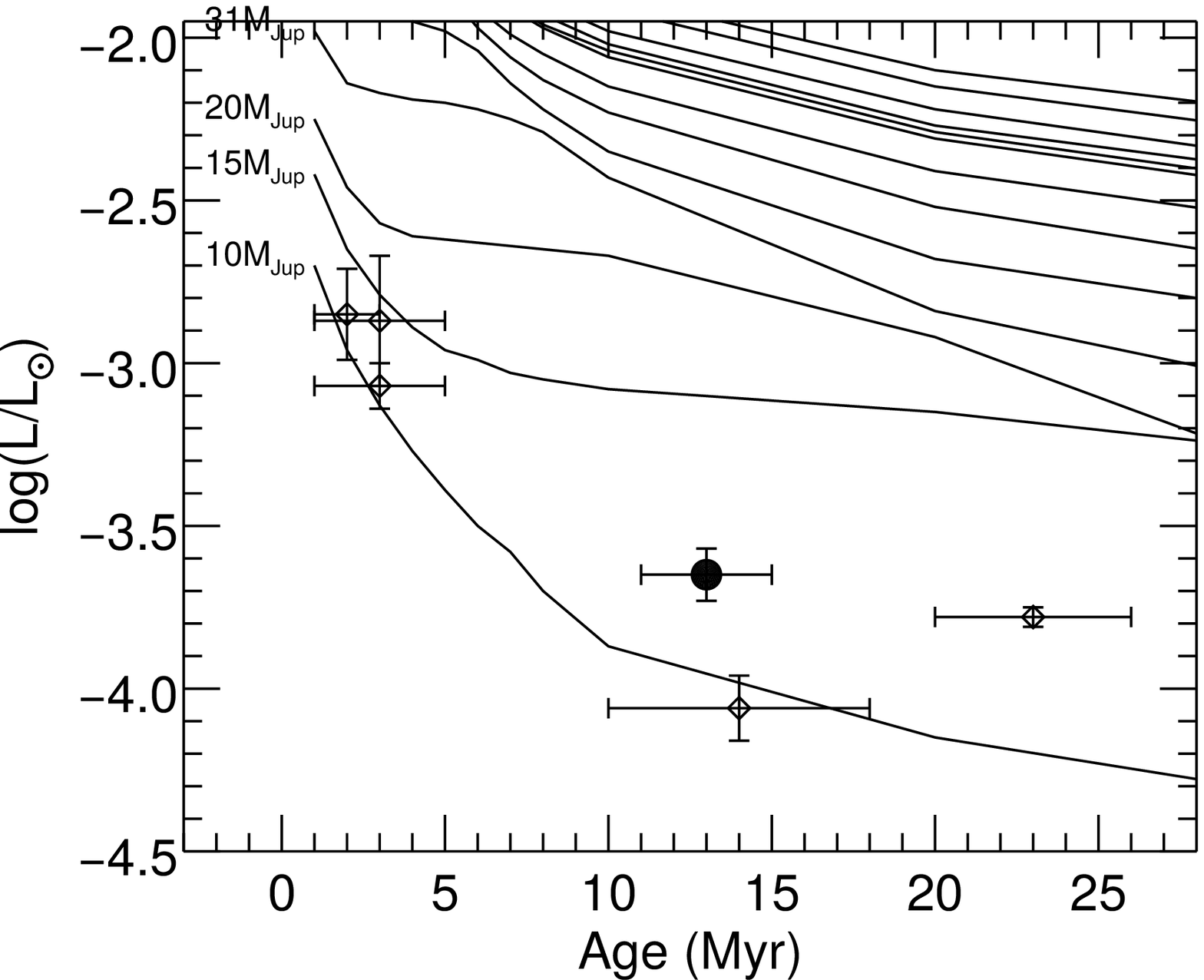}}
    \put(0.88,1.15){\fontfamily{phv}\selectfont BT-SETTL}
    \put(0.00,0.40){\includegraphics[width=\columnwidth]{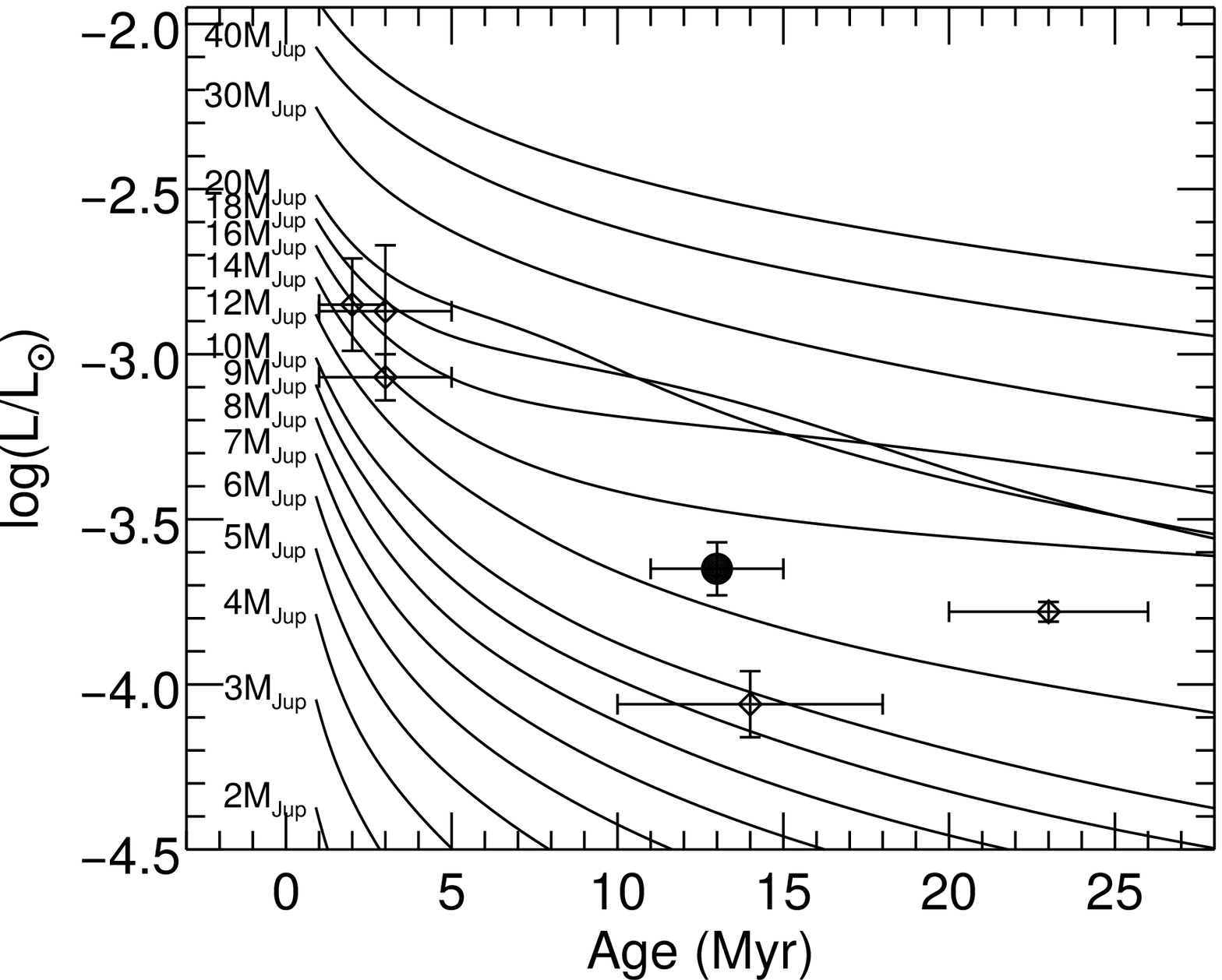}}
    \put(0.36,0.75){\fontfamily{phv}\selectfont BEX Hottest}
    \put(0.50,0.40){\includegraphics[width=\columnwidth]{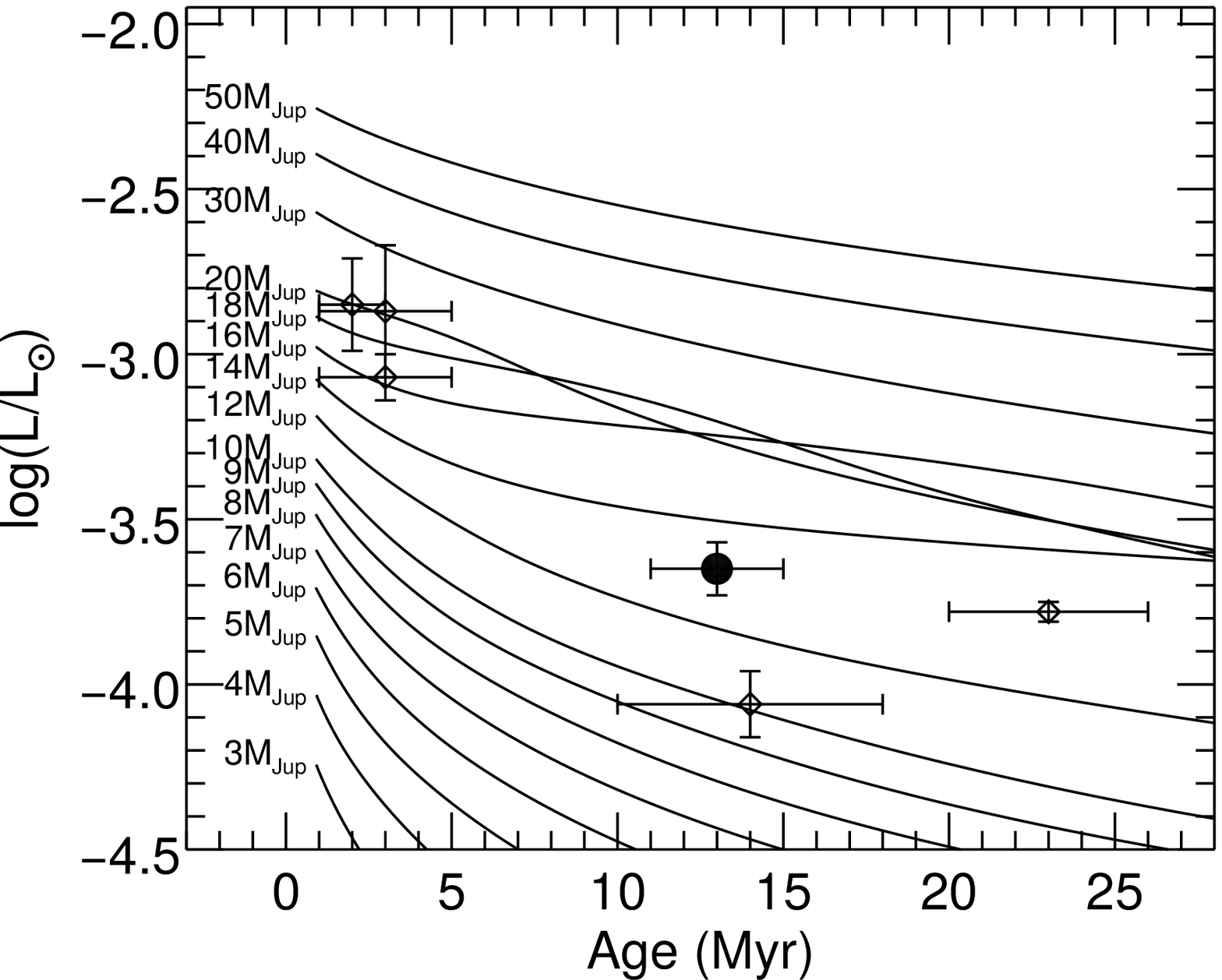}}
    \put(0.895,0.75){\fontfamily{phv}\selectfont BEX Hot}
    \put(0.00,0.00){\includegraphics[width=\columnwidth]{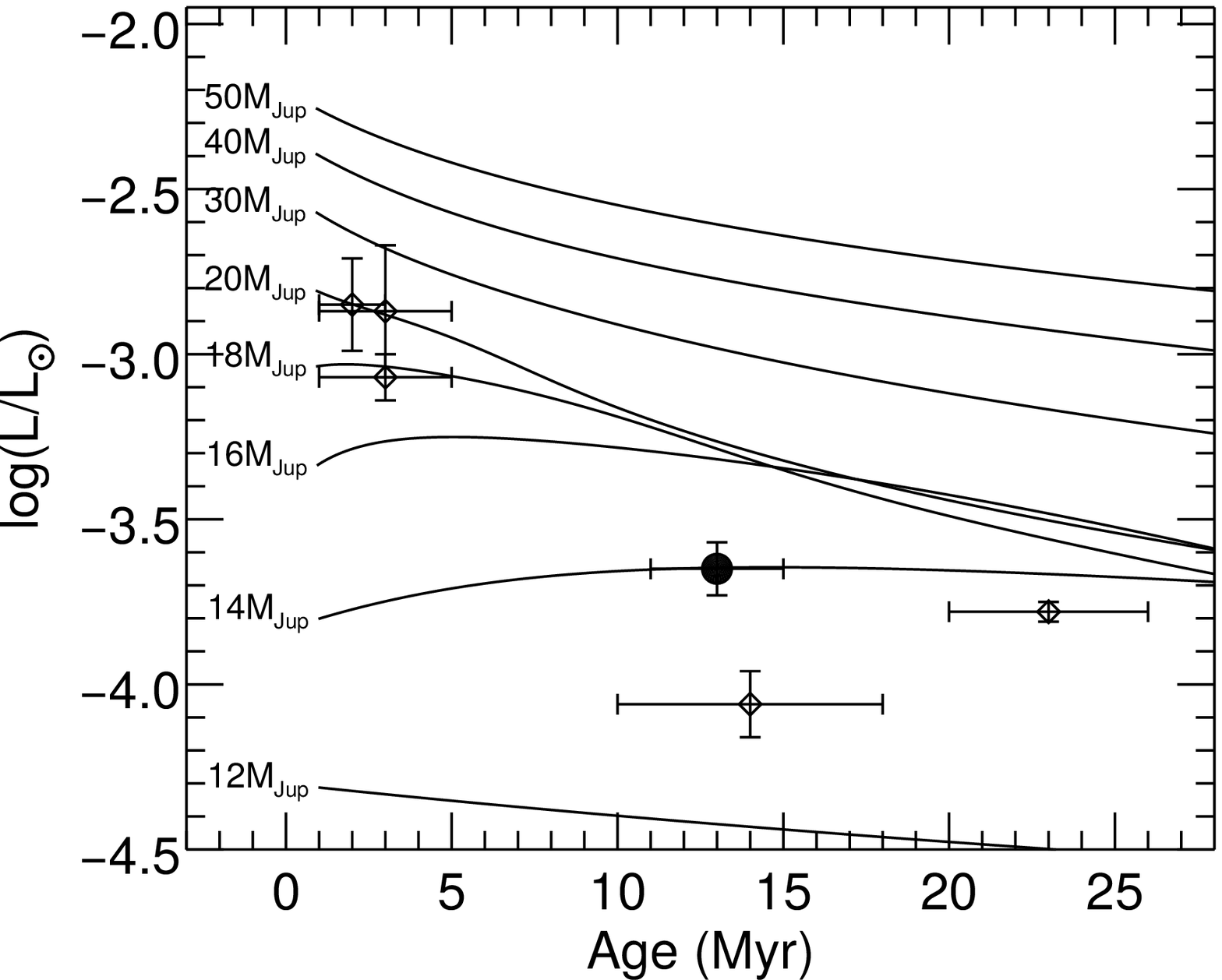}}
    \put(0.38,0.35){\fontfamily{phv}\selectfont BEX Cold}
\end{picture}
\caption{\label{fig:mass} \hdb's luminosity and age (filled circle) compared to evolutionary tracks. References and extracted mass estimates are listed in Table~\ref{tab:mass}. The locations of other young, low-mass companions are shown for comparison \citep[open symbols, target names in the first panel;][]{cha17,bow16}.}
\end{figure*}

Fig.~\ref{fig:mass} shows \hdb\ with respect to the evolutionary tracks. 
We adopt an age of 13$\pm$2\,Myr, assuming that there is no significant delay of the formation of \hdb\ compared to other members of Upper Scorpius and its host star. To estimate the mass of \hdb, we linearly interpolate each grid at the age and luminosity of \hdb. Mass estimates, listed in Table~\ref{tab:mass}, range from $M$\,=\,12.5$\pm$1.0\,\MJup\ to $M$\,=\,14.2$^{+0.4}_{-0.9}$\,\MJup, depending on the initial entropy of the formation scenario, i.e., whether a hot star or cold start model was used, and on the atmospheric dust model. 
When assuming a 3\,Myr formation timescale \citep[cf.][]{for05} for \hdb\ (i.e., an effective age of 10$\pm$2\,Myr), the derived masses decrease by 0--0.7\,\MJup\ (see Table~\ref{tab:mass}).

\begin{table}
  \caption{Masses from isochrone fits\label{tab:mass}}
\centering
\begin{tabular}{lccc}
\hline\hline
Model & Ref. & $M$ (\MJup) & $M_{t_{\rm form}=3\,{\rm Myr}}$ (\MJup) \\
\hline
DUSTY                          & 1 & $12.3^{+0.9}_{-0.8}$ & $11.9^{+2.5}_{-0.9}$ \\
BT-SETTL                       & 2 & $12.3^{+0.8}_{-0.7}$ & $11.9^{+1.7}_{-0.8}$ \\
BEX Hottest                    & 3 & $12.8^{+1.1}_{-0.8}$ & $12.1^{+1.7}_{-1.1}$ \\
BEX Hot                        & 3 & $13.1^{+0.8}_{-0.6}$ & $12.6^{+1.3}_{-0.9}$ \\
BEX Cold/Warm\tablefootmark{a} & 3 & $14.0^{+0.2}_{-0.5}$ & $14.0^{+0.2}_{-0.4}$ \\
\hline
\end{tabular}
\tablefoot{
  \tablefoottext{a}{The cold and warm evolutionary tracks of the BEX models predict identical luminosities for objects above $\sim$10\,\MJup\ \citep{mor17}.}
}
\tablebib{(1) \citealt{cha00}, (2) \citealt{bar15}, (3) \citealt{mor17}.}
\end{table}

\subsection{Upper limits on accretion}
We do not detect any evidence for line emission, in particular not in the accretion-sensitive hydrogen features, i.e., Paschen-$\beta$ (Pa$\beta$, 1.282\,\mum) and Brackett-$\gamma$ (Br$\gamma$, 2.167\,\mum), or in the limit of the Brackett series ($H$ band). Accordingly, we do not see evidence for strong accretion activity onto \hdb. Assuming magnetospheric accretion from an isolated circumplanetary disk onto \hdb,\ we infer a quantitative upper limit for the mass accretion rate based on the upper limit for the accretion luminosity $L_\mathrm{acc}$ and empirical $L_\mathrm{acc}$--$L_\mathrm{line}$ relations derived for T~Tauri stars \citep{alc14}.

We use the description by \citet{gul98} to estimate a mass accretion rate from an accretion luminosity using
\begin{equation}\label{eq:gul98}
  \dot{M} = \left(1-\frac{R_{\rm pl}}{R_\mathrm{in}}\right)^{-1}\frac{L_\mathrm{acc}\,R_{\rm pl}}{G\,M_{\rm pl}}\quad.
\end{equation}
Following \citet{lov11}, we conservatively set the inner disk truncation radius $R_\mathrm{in}$=2.7\,$R_{\rm pl}$. Larger radii, such as $R_\mathrm{in}$=5\,$R_{\rm pl}$, which are commonly used for T Tauri stars, result in stronger constraints on the inferred upper limit of the mass accretion rate of up to a factor of 1.5. The mass and radius of the companion were set to $M_{\rm pl}$=12.3\,$\pm$\,0.8\,\MJup\ and $R_{\rm pl}$=1.6$\,R_\mathrm{Jup}$, which is consistent with our derived luminosity and temperature. Flux calibration of the spectra was achieved by normalizing the spectra, after multiplication with the spectral response curves of the respective filters, to the $J$ and \Ks\ magnitudes measured by \citet{bai14}\footnote{No $H$-band magnitudes have been published so far, so this technique could not be applied to our $H$-band spectrum.}. The measured continuum noise values in Pa$\beta$ and Br$\gamma$ of $\Delta F {\rm d}\lambda$(Pa$\beta$)\,=\,1.4$\times$$10^{-18}$\,erg/s/cm$^2$ and $\Delta F {\rm d}\lambda$(Br$\gamma$)\,=\,1.6$\times$$10^{-18}$\,erg/s/cm$^2$ translate to upper limits of the line luminosity $L_\mathrm{line}$\,=\,$4\pi F_\mathrm{line}d^2$ of $\log(L_{\rm line}/L_\sun)$\,$<$\,$-8.78$ and $<$\,$-8.95$, respectively (99\% confidence). Using the $L_\mathrm{acc}$--$L_\mathrm{line}$ correlations listed \citet{alc14} we infer $\log(L_\mathrm{acc}/L_\sun)$\,<\,$-5.63$ (Pa$\beta$) and $\log(L_\mathrm{acc}/L_\sun)$\,<\,$-6.20$ (Br$\gamma$). Using eq.~(\ref{eq:gul98}) we infer upper limits for the mass accretion rate of $\dot{M}<1.8\times10^{-9}$\,\MJup{}yr$^{-1}$ (Pa$\beta$) and $\dot{M}<4.8\times10^{-10}$\,\MJup{}yr$^{-1}$ (Br$\gamma$).

Uncertainties and upper limits were calculated using a Monte Carlo simulation taking into account the uncertainties in measured flux, mass, distance, and the empirical $L_\mathrm{acc}$--$L_\mathrm{line}$ relations \citep{alc14}. We conservatively account for systematic uncertainties such as the flux calibration and the subtraction of hydrogen absorption in the telluric standard star spectra by increasing the spectral noise by a factor of 2. All quoted upper limits are 99\% confidence limits. 
Since the mass accretion rate as calculated from Br$\gamma$ results in the more restrictive limit, we quote $\dot{M}<4.8\times10^{-10}$\,\MJup{}yr$^{-1}$ as our best estimate of the mass accretion rate for \hdb.

We make a prediction for the expected accretion-induced line luminosity at the wavelength of H$\alpha$ based on the derived line luminosity (using the more restrictive limit $\log(L_\mathrm{acc}/L_\sun)$\,<\,$-6.20$~). We use the $L_\mathrm{acc}$--$L_\mathrm{line}$ relation from \citet{alc14} to derive $\log(L_{\rm H\alpha}/L_\sun)$\,$<$\,$-6.69$ (99\% confidence).

\section{Summary and discussion}\label{sec:summary}
We present new high S/N (20--50/pix) intermediate-resolution VLT/SINFONI 1--2.5\,\mum\ integral field spectroscopy of \hdb, which is a companion at the deuterium burning limit in the Upper Scorpius association.
We detect a number of spectral features indicative of a low mass for \hdb, and also the H-band continuum shows the typical triangular shape of low gravity objects.
Comparison with spectral libraries of young and field objects as well as analyzing spectral features suggests a spectral type of L1.5$\pm$1.0. Previous estimates based on low-resolution spectroscopy classified this object as either L2 (based on $H$-band spectroscopy) or L3 \citep[$K$-band spectroscopy;][]{bai14}. 

We use this new information and the recently published distance of \hdb\ \citep[102.8$\pm$2.5\,pc]{gai16} to derive a luminosity of $\log\left(L_\mathrm{bol}/L_\odot\right)=-3.65\pm0.08$ and an effective temperature of \Teff\,=\,1820\,$\pm$\,240\,K;  this recently published distance of this binary star is $\sim$10\% larger than the previous best estimate from the Hipparcos mission. Based on its luminosity and age compared to predictions from evolutionary models, we estimate a mass between $M$\,=\,12.3$^{+0.8}_{-0.7}$\,\MJup\ (hot start) and $M$\,=\,14.0$^{+0.2}_{-0.5}$\,\MJup\ (cold start) in the limit of zero formation time. Mass predictions are slightly lower if a formation timescale of 3\,Myr is assumed of between $M$\,=\,$11.9^{+2.5}_{-0.9}$\,\MJup\ and $M$\,=\,14.0$^{+0.2}_{-0.4}$\,\MJup.

The formation and early evolution of \hdb\ remain unclear. The low mass ratio of $q$$\sim$0.004 with respect to the host system \citep[$M_{\rm HD106906AB}$\,=\,1.37\,\Msun+1.34\,\Msun;][]{lag17} suggests ``planet-like'' formation in a protoplanetary disk \citep{pep14,reg16}. Formation through core accretion, however, is restricted to orbits much closer to the star than its current separation of $>$700\,AU \citep[e.g., $\lesssim$35\,AU in the simulations by][]{dod09}. Accordingly, subsequent migration would be required. Currently, however, no conclusive evidence for scattering with an internal planet or the central binary has been found \citep[cf.][]{jil15}. An ejection scenario with subsequent braking through a stellar fly-by appears unlikely in the light of the simulations by \citet{rod17}. While not strictly excluded, in situ formation through disk instability appears unlikely (though not excluded) because large disks $>$700\,AU have been rarely observed around forming stars. Formation of \hdb\ in a star-like channel, i.e., through direct collapse of the parent cloud core into a triple system \hd{}AB+b, remains a possibility.

The degeneracy between the various formation scenarios could be broken when the orbit of \hdb\ around its host system becomes known. Its long orbital period of $>$3000\,yr \citep{jil15}, however, precludes precise direct measurement of the motion of \hdb. Indirect information on the orbit of \hdb, however, may be gathered from observations of the circumstellar disk whose eccentricity is thought to reflect interaction with the companion \citep{nes17}. Alternatively, additional high precision spectroscopy in the atmospheric absorption bands (e.g., near 1.4\,\mum\ and 2\,\mum) and at mid-IR wavelengths together with retrieval analysis \citep[e.g.,][]{lin12} may be used to measure the chemical composition of the atmosphere of \hdb,  which is a function of its birth location. For example, an enhanced carbon-to-oxygen ratio suggests a formation in a disk close to the host star \citep[e.g.,][]{obe11}. Direct collapse, in contrast, would produce a stellar abundance ratio. Future observation with high-precision spectroscopy instruments such as those on board the James Webb Space Telescope will help to break these degeneracies.

While the presented spectra cover a number of hydrogen features (Paschen-$\beta$, Brackett-$\gamma,$ and higher), none of these features appears in emission. We thus exclude magnetospheric accretion onto \hdb\ at a rate of $\dot{M}>4.8\times10^{-10}$\,\MJup{}yr$^{-1}$ (99\% confidence). With this measurement we can exclude accretion at a level several times below the accretion levels that have been observed for other targets at a similar mass and age as \hdb. For example, \citep{joe13} find a mass accretion rate of $\dot{M}=8\times10^{-9}$\,\MJup{}yr$^{-1}$ for OTS\,44, a free-floating 12\,\MJup\ member of the $\sim$2\,Myr-old Chamaeleon\,I region. DH\,Tau, an 11\,\MJup\ companion at $\sim$340\,AU from its $\sim$1--2\,Myr-old parent star, is thought to accrete at a rate of $\dot{M}=3.3\times10^{-9}$\,\MJup{}yr$^{-1}$ \citep{zho14}. For comparison with optical studies and as a reference for optical follow up, we predict an accretion-induced H$\alpha$ line luminosity of $\log(L_{\rm H\alpha}/L_\sun)$\,$<$\,$-6.69$. Previous observations of substellar companion candidates in H$\alpha$ returned values of $\log(L_{\rm H\alpha}/L_\sun)$\,$=$\,$-4.2$ \citep[LkCa\,15b;][]{sal15} and $\log(L_{\rm H\alpha}/L_\sun)$\,$=$\,$-3.3$ \citep[HD\,142527B;][]{clo14}, again significantly larger than our limit for \hdb. If \hdb\ were accreting, its rate must be significantly below that of these comparison objects. We thus conclude that \hdb\ features no significant amount of circumplanetary gas. However, the present data do not exclude the presence of a \emph{gas-poor} disk around \hdb, similar to what has been observed for the primary object \hd\ \citep{kal15,lag16}. In fact, \citet{kal15} present weak evidence for the presence of circumplanetary dust around \hdb\ based on color excesses and a marginally spatially extended image of the source. 

\begin{figure*}[tbh]
  \centering
  \setlength{\unitlength}{\textwidth}
\begin{picture}(1.0,0.7)
    \put(0.00,0.00){\includegraphics[width=1\textwidth]{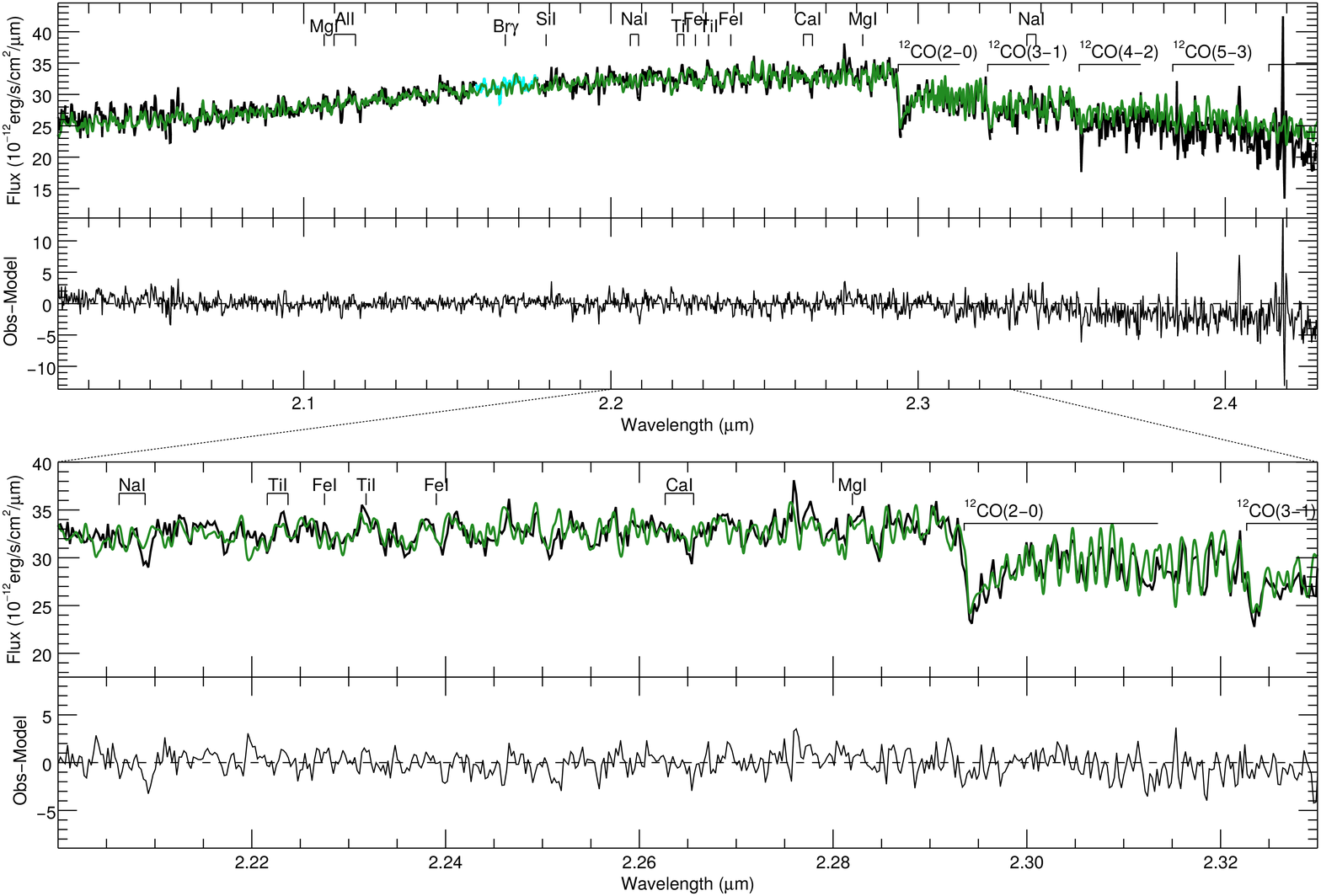}}
    \put(0.06,0.65){\fontfamily{phv}\Large\selectfont a)}
    \put(0.06,0.48){\fontfamily{phv}\Large\selectfont b)}
    \put(0.06,0.3){\fontfamily{phv}\Large\selectfont c)}
    \put(0.06,0.135){\fontfamily{phv}\Large\selectfont d)}
\end{picture}
\caption{\label{fig:models1}{\bf  (a)} \Teff=1600\,K, \logg=3.5 BT-SETTL model \citep[green curve;][]{bar15}, smoothed to the resolution of our K-band spectrum \hdb\ (black) and scaled to the same average flux. Most features observed in the spectra are reproduced by the (virtually noise-free) model, supporting evidence that the measured features are caused by atmospheric absoprtion rather than noise. {\bf  (b)} Residuals are shown.\  {\bf (c)} and {\bf (d)} show the same for a shorter wavelength range.}
\end{figure*}

Our new study makes \hdb\ one of the few low-\Teff\ companions for which high S/N spectroscopy could be observed. As demonstrated in Fig.~\ref{fig:models1}, most of the observed features in the spectrum must be real rather than due to measurement uncertainties\footnote{We note that the selected model temperature is slightly below our best estimate of \Teff=1820\,K. This selection was made due to a bimodal $\chi^2$ distribution of BT-SETTL model fits in \logg--\Teff\ space which has local minima at \Teff$\approx$1600\,K and \Teff$\sim$2600\,K with a particularly poor representation of the K-band spectrum of \hdb\  by \Teff$\approx$1700--2100\,K models. Detailed model fitting to explore reasons for this bimodality is beyond the scope of the current paper.}.
These high fidelity observations were possible with current instrumentation because of the large angular separation of \hdb\  from its host star. Most other directly imaged systems feature much smaller separations, which reduce the achievable S/N due to additional light from the primary. The current spectrum thus serves as a reference spectrum for future studies of young low-gravity objects in orbit around low-mass stars, in particular with regard to the higher sensitivity of the James Webb Space Telescope and the higher angular resolution of the upcoming generation of 30 m-class telescopes.

\begin{acknowledgements}
  We thank the anonymous referee for a thorough review and valuable suggestions for improvement.
  We thank Michael Line for valuable contributions in preparation of the observing time proposal. 
  This work has been carried out within the frame of the National Centre for Competence in Research PlanetS supported by the Swiss National Science Foundation. S.D., S.P.Q., and G.-D.M.\ acknowledge the financial support of the SNSF. C.M. and G.-D.M. acknowledge the support from the Swiss National Science Foundation under grant BSSGI0$\_$155816 ``PlanetsInTime''. 
  This work has made use of data from the European Space Agency (ESA) mission {\it Gaia} (\url{http://www.cosmos.esa.int/gaia}), processed by the {\it Gaia} Data Processing and Analysis Consortium (DPAC; \url{http://www.cosmos.esa.int/web/gaia/dpac/consortium}). Funding for the DPAC has been provided by national institutions, in particular the institutions participating in the {\it Gaia} Multilateral Agreement.
\end{acknowledgements}

\end{document}